\documentclass[aps,amsmath,amssymb,nofootinbib]{revtex4}
\usepackage{graphicx,subfigure}
\def\beq{\begin{equation}}
\def\eeq{\end{equation}}
\def\beqr{\begin{eqnarray}}
\def\eeqr{\end{eqnarray}}
\def\bdpm{\begin{displaymath}}
\def\edpm{\end{displaymath}}
\def\npb#1#2#3 {Nucl. Phys. B {\bf#1}, #2 (#3)}
\def\plb#1#2#3 {Phys. Lett. B {\bf#1}, #2 (#3)}
\def\prd#1#2#3 {Phys. Rev. D {\bf#1}, #2 (#3)}
\def\jhep#1#2#3 {J. High Energy Phys. {\bf#1}, #2 (#3)}
\def\jpg#1#2#3 {J. Phys. G {\bf#1}, #2 (#3)}
\def\epj#1#2#3 {Eur. Phys. J. C {\bf#1}, #2 (#3)}
\def\arnps#1#2#3 {Ann. Rev. Nucl. Part. Sci. {\bf#1}, #2 (#3)}
\def\ibid#1#2#3 {{\it ibid.} {\bf#1}, #2 (#3)}
\def\none#1#2#3 {{\bf#1}, #2 (#3)}
\def\mpla#1#2#3 {Mod. Phys. Lett. A {\bf#1}, #2 (#3)}
\def\pr#1#2#3 {Phys. Rep. {\bf#1}, #2 (#3)}
\def\prl#1#2#3 {Phys. Rev. Lett. {\bf#1}, #2 (#3)}
\def\ptp#1#2#3 {Prog. Theor. Phys. {\bf#1}, #2 (#3)}
\def\rmp#1#2#3 {Rev. Mod. Phys. {\bf#1}, #2 (#3)}
\def\zpc#1#2#3 {Z. Phys. C {\bf#1}, #2 (#3)}

\begin{document}

\begin{flushright}
KIAS-P09041
\end{flushright}
\vspace{0.7in}
\title{\Large \bf $SU(4)_L \times U(1)_X$ models with little Higgs}
\date{\today}

\author{ \bf Soo-hyeon Nam$^{a,b}$\footnote{Email:glvnsh@gmail.com},
Kang Young Lee$^{c}$\footnote{Email:kylee@muon.kaist.ac.kr},
and Yong-Yeon Keum$^{b,d}$\footnote{Email:yykeum@gmail.com}
 }

\affiliation{ $^{a}$ Korea Institute of Science and Technology Information, Daejeon 305-806, Korea \\
$^{b}$ School of Physics, KIAS, Seoul 130-722, Korea \\
$^{c}$ Division of Quantum Phases \& Devices, School of Physics, Konkuk University, Seoul 143-701, Korea   \\
$^{d}$ Department of Physics and BK21 Initiative for Global Leaders in Physics, Korea University, Seoul 136-701, Korea
}

\begin{abstract}
We discuss the aspects of the fermions and gauge bosons 
in $SU(4)_L \times U(1)_X$ models with little Higgs. 
We introduce a set of fermions which ensures the cancellation of gauge anomaly, 
and explicitly show the cancellation of one-loop quadratic divergence 
to the Higgs mass from all fermion multiplets and all gauge bosons. 
We present the interactions of the standard model fermions 
with the physical gauge bosons. 
We also discuss some phenomenological implications of the model 
based on recent experimental results.
\end{abstract}

\maketitle

\section{Introduction}

 An extension of the standard model (SM) gauge group $SU(2)_L\times U(1)_Y$ to $SU(4)_L\times U(1)_X$ has been proposed by various authors due to its several distinctive features.  For instance, electroweak unification could be obtained in an $SU(4)_L$ model with its subgroup $SU(2)_L \times SU(2)_R\times U(1)$ and $\sin^2\theta_W = 1/4$ in the left-right symmetry limit \cite{Fayyazuddin84}.  Also, one can construct a $SU(4)_L$ model in the lepton sector in which the  hypothetical large neutrino magnetic moment around $10^{-11}$ of the Bohr magneton is naturally compatible with acceptably small neutrino mass of a few $eV$ \cite{Voloshin88}.  Most interestingly, the gauged $SU(4)_L\times U(1)_X$ group including both quarks and leptons can provide an answer to the question why we only observe three families of fermions in nature, in a sense that anomaly cancellation is achieved when $N_f = N_c = 3$ where $N_f (N_c)$ is the number of families (colors) \cite{Pleitez93}.  A systematic way of constructing anomaly-free fermion spectra with $SU(4)_L\times U(1)_X$ gauge group has been discussed in Ref. \cite{Ponce07}.

 Instead of the usual Higgs mechanism, recently, the little Higgs mechanism has been implemented in $SU(4)_L\times U(1)_X$ gauge group by Kaplan and Schmaltz (K-S) as an alternative solution to the hierarchy and fine-tuning issues \cite{Kaplan03}. Little Higgs models (LHMs) adopts the early idea that Higgs can be considered as a Nambu Goldstone boson from global symmetry breaking at some higher scale $\Lambda \sim 4\pi f$ \cite{Dimopoulos82} and acquires a mass radiatively through symmetry breaking at the electroweak scale $v$ by collective breaking \cite{Arkani01}.  The LHM with the $SU(4)_L\times U(1)_X$ gauge symmetry appears fundamentally different from other types of LHMs due to the multiple breaking of global symmetry by separate scalar fields \cite{Schmaltz05}.  The bound on the new symmetry breaking scale $f$ of this K-S model was obtained earlier from the tree-level electroweak constraints in Ref. \cite{Csaki03}.

 The novel feature of LHMs is that Higgs mass is protected by a global symmetry which is spontaneously broken and so the one-loop quadratic divergences to the Higgs mass are canceled by particles of the \textsl{same} spin; i.e., a new fermion cancels a quadratic divergence from a SM fermion.  The K-S model has vector-like heavy quarks of charge 2/3 for each generation with a simple family universal embedding, but this choice leaves nonvanishing quadratic divergences from light fermions and gauge anomalies which require additional fermion multiplets at the scale $\Lambda$. Instead, we introduce alternative fermion set with which the one-loop quadratic divergence to the Higgs mass and gauge anomaly are absent at the scale $f$.  In our case, the cancellation of quadratic divergences is made between a SM particle and its new heavy partner not only of the \textsl{same} spin but also of the \textsl{same} electric charge so that the one-loop quadratic divergences from all fermion loops are canceled. We will explicitly show this cancellation mechanism in the next section.

 In general $SU(4)_L\times U(1)_X$ models, anomaly cancellation can be achieved  by embedding the first two generations of quarks into $\bar{4}$ representations of $SU(4)_L$ while the third generation of quarks and all three generations of leptons are into $4$s of $SU(4)_L$ \cite{Ponce07}.  The idea of adopting this alternative choice of fermion set to the LHM with simple group was first proposed in Ref. \cite{Kong03}, and a similar study on the K-S model with $SU(3)_L\times U(1)_X$ gauge group can be found in Ref. \cite{Schmaltz06,Han06}.  However, in order to have a proper Higgs mass, the $SU(3)$ model requires so-called $\mu$-term that manifestly breaks the global $SU(3)^2$ symmetry and gives tree-level masses to scalar particles including Higgs, which ruins the original motivation of the LHM to acquire Higgs mass spontaneously \cite{Kaplan03}.  In this paper, we modify the K-S model with $SU(4)_L\times U(1)_X$ gauge symmetry by embedding anomaly-free sets of fermions in Sec. II, and discuss the phenomenological implications of this model in Sec III in further detail.

\section{Models}

\subsection{Scalar Sector}

The scalar sector in $SU(4)_L \times U(1)_X$ models with little Higgs is based on the non-linear sigma model describing $[SU(4)/SU(3)]^4$ global symmetry breaking with the diagonal $SU(4)$ subgroup gauged and four non-linear sigma model field $\Phi_i$ where $i = 1,2,3,4$.
The standard $SU(2)_L\times U(1)_Y$ gauge group can be embedded into the theory with an additional $U(1)_X$ group.  The $SU(4)$ breaking is not aligned and only the gauged $SU(2)$ is linearly realized in this model where the scalar fields $\Phi_i$ can be parametrized as
\beq
\Phi_{1}=e^{+i {\cal H}_u \frac{f_{2}}{f_{1}}}
        \left( \begin{array}{l} 0  \\ 0 \\ f_{1} \\0 \end{array} \right) ,\quad
\Phi_{2}=e^{- i {\cal H}_u \frac{f_{1}}{f_{2}}}
        \left( \begin{array}{l} 0  \\ 0 \\ f_{2} \\0 \end{array} \right) ,
        \nonumber
\eeq
\beq \label{Eq:Scalarfield}
\Phi_{3}=e^{+i {\cal H}_d \frac{f_{4}}{f_{3}}}
        \left( \begin{array}{l} 0  \\ 0 \\ 0 \\f_{3} \end{array} \right) ,\quad
\Phi_{4}=e^{- i {\cal H}_d \frac{f_{3}}{f_{4}}}
        \left( \begin{array}{l} 0  \\ 0 \\ 0 \\f_{4} \end{array} \right) ,
\eeq
where
\beq \label{Eq:Higgs}
  {\cal H}_u =\left[\Pi_u +
    \left( \begin{array}{ccc}
           \begin{array}{cc} 0 & 0 \\ 0 & 0 \end{array}
             & h_u & \begin{array}{c} 0 \\ 0 \end{array}   \\
            h_u^\dagger & 0 &  0 \\
           \begin{array}{cc}  0 &  0 \end{array} & 0 & 0 \\
           \end{array} \right)\right]\Big{/}f_{12} ,
\quad
  {\cal H}_d =\left[\Pi_d +
    \left( \begin{array}{ccc}
      \begin{array}{cc} 0 & 0 \\ 0 & 0 \end{array}
      &  \begin{array}{c}  0 \\  0 \end{array}  &  h_d \\
        \begin{array}{cc}  0 &  0 \end{array}  & 0 &  0 \\
      h_d^\dagger &  0 & 0 \\
    \end{array} \right)\right]\Big{/}f_{34} ,
\eeq
and $f_{ij} = \sqrt{f^2_{i} + f^2_{j}}$.  Here we only show the two complex Higgs doublets $h_{u,d}$ and discard the other singlets in $\Pi_q\ (q=u, d)$ whose contributions to fermion and gauge boson masses are negligible.  The detailed discussion on this scalar sector including the Higgs potential can be found in Ref. \cite{Kaplan03}, so we do not repeat it here and lead the readers to the original paper.  Instead, in this paper, we mainly focus on the fermion and gauge boson sectors.

\subsection{Fermion Sector}

 The covariant derivative of the scalar and fermion quadruplets is given by
\beq \label{eq:CovariantD}
D^\mu=\partial^\mu + ig T_\alpha A^\mu_\alpha + ig_X X A_X^\mu ,
\eeq
where $A^\mu_\alpha, g$ and $A_X^\mu, g_X$ are the gauge bosons and couplings of the $SU(4)_L$ and $U(1)_X$ gauge groups, respectively, and $T_\alpha = \lambda_\alpha/2$ with $\lambda_\alpha$ the Gell-Mann matrices for $SU(4)_L$ normalized as Tr$(\lambda_\alpha\lambda_\beta) = 2\delta_{\alpha\beta}$.
The electric charge generator is a linear combination of the $U(1)_X$ generator and the three diagonal generators of the $SU(4)_L$ group:
\beq
Q = a_1T_{3}+\frac{a_2}{\sqrt{3}}T_{8}+ \frac{a_3}{\sqrt{6}}T_{15}+ XI_4 ,
\eeq
with
\beq
T_3 = \frac{1}{2}\textrm{diag}(1,-1,0,0),\quad T_8 = \frac{1}{2\sqrt{3}}\textrm{diag}(1,1,-2,0), \nonumber
\eeq
\beq
T_{15} = \frac{1}{2\sqrt{6}}\textrm{diag}(1,1,1,-3), \quad I_4 = \textrm{diag}(1,1,1,1),
\eeq
where $X$ is the $U(1)_X$ charge.  The free parameters $a_1$, $a_2$ and $a_3$ fix the electroweak charges of the gauge, scalar, and fermion representations.  In particular, $a_1=1$ gives the correct embedding of the SM isospin $SU(2)_L$ doublets ($Q = T_3 + Y$),
and the remaining parameters $a_2$ and $a_3$ can be uniquely determined once the electric charges of the fermion multiplets are specified.

Since this LHM has a gauged $SU(4)_L$, the SM doublets must be expanded to $SU(4)_L$ quadruplets, and the extra fermions in the quadruplets should cancel the quadratic divergence from the SM fermion (especially from the top).  Taking into account this requirement, one can embed the SM doublet ($t,b$) into following two types of
$SU(4)_L$ quadruplet $\psi_L$:
\beq
\psi_L^{I} = (t, b, T, T')_L^T , \quad \psi_L^{II} = (t, b, T, B)_L^T ,
\eeq
or their charge conjugates \cite{Kong03}.
Here we choose $T$ and $T'(B)$ to have the same electric charge as $t(b)$, and
the duplicated extra heavy fermions $T$ and $T'(B)$ remove the quadratic divergences due to their SM fermion partners $t(b)$. This choice also avoids other fermions and bosons having exotic fractional electric charges \cite{Ponce07}. In this case, the corresponding hypercharge generator $Y$ becomes \footnote{$T_{8}'$ and $T_{15}'$ in this paper are defined as $T_{12}$ and $T_{15}$, respectively, in Ref \cite{Kaplan03}.}
\beq
Y^{I} =  \frac{1}{\sqrt{2}}T_{15}' + XI_4, \quad Y^{II} =  T_{8}' + XI_4 ,
\eeq
where
\beq
T_{8}' = -\frac{1}{\sqrt{3}}T_{8}- \frac{1}{\sqrt{6}}T_{15} , \quad
T_{15}' = -\sqrt{\frac{2}{3}}T_{8} + \frac{2}{\sqrt{3}}T_{15} .
\eeq
When the global $SU(4)$ symmetry is broken down to the global $SU(3)$ symmetry, the diagonal generator $T_{15}'(T_8')$ is broken and $T_8'(T_{15}')$ is left unbroken for Type I(II).
Meanwhile, the local $SU(4)_L \times U(1)_X$ symmetry is broken down to the local $SU(2)_L \times U(1)_Y$ symmetry, and the mixed diagonal generators $T_8$, $T_{15}$, and $I_4$ result in the unbroken hypercharge generator $Y$ and two broken generators $\eta_u$ and $\eta_d$.
After symmetry breaking, $\eta_u$ and $\eta_d$ are realized as two pseudo-scalar singlets residing in the fields $\Pi_u$ and $\Pi_d$, respectively. Similar $\eta$ singlets appear in various LHMs, and their collider signatures are reviewed in Ref \cite{Kilian05}.

In the original K-S model, the quark quadruplets belong to Type I, and all three generations have identical quantum numbers (universal embedding). As mentioned earlier, such a universal fermion sector requires additional ambiguous fermion multiplets at the scale $\Lambda$ in order to remove the gauge anomaly in the UV completion of the model.  Alternatively, one can make the LHMs (including the K-S model) anomaly-free by taking different charge assignments for the different generations of quark multiplets (anomaly-free embedding).  Since the Type I case in the LHM was discussed already in Refs \cite{Kaplan03,Csaki03} (although it is anomalous), we mainly focus on the Type II case in this paper.

One benefit of embedding Type II (over Type I) quark quadruplets in the LHM is that the cancellation of the quadratic divergences is even made between a light SM fermion and its heavy fermion partner of the `same' electric charge. In the case of the universal embedding, the quarks and the leptons of each generation are put into \textbf{4} representation of $SU(4)_L$:
{\setlength\arraycolsep{2pt} \beqr \label{Eq:fermions}
\psi_m^{q} &=& (u, d, iU, iD)_m^T , \qquad iu_m^c, id_m^c, iU_m^c, iD_m^c ,\cr
\psi_m^{\ell} &=& (\nu, e, iN, iE)_m^T,  \qquad ie_m^c, iE_m^c ,
\eeqr }
where $m$ is the generation index, and the superscripts $q$ and $\ell$ denote quark and lepton states, respectively.  In addition to the above states, one can add Weyl singlet states with no X charges without spoiling the anomaly constraint.  As discussed earlier, the anomaly-free choice of fermion set in this model is uniquely obtained by putting the first two generations of quarks into $\bar{\bf 4}$ representations of $SU(4)_L$ while the rest of the quark and lepton families are the same as those in Eq. (\ref{Eq:fermions}):
{\setlength\arraycolsep{2pt} \beqr
\psi_1^{q} &=& (d, u, iD, iU)^T , \qquad id^c, iu^c, iD^c, iU^c ,\cr
\psi_2^{q} &=& (s, c, iS, iC)^T , \qquad is^c, ic^c, iS^c, iC^c .
\eeqr }
Then, in the case of Type II, the $X$ charges of the scalar and fermion multiplets are given as
\beq
X(\Phi_{1,2}) = -\frac{1}{2},\quad X(\Phi_{3,4}) = \frac{1}{2},\quad X(\psi^q_L) = \frac{1}{6},\quad X(\psi^\ell_L) = -\frac{1}{2} ,\quad X(\psi_R) = \mathnormal{Q} ,
\eeq
and the two doublet Higgs fields $h_{u,d}$ shown in Eq. (\ref{Eq:Higgs}) are of the following form
\beq \label{Eq:Higgsfield}
h_u = \frac{1}{\sqrt{2}}{H_u^0 \choose H_u^-} , \quad  h_d = \frac{1}{\sqrt{2}}{H_d^+ \choose H_d^0},
\eeq
where the neutral components of the two Higgs fields develop vacuum expectation values (vevs):
\beq \label{Eq:Higgsvev}
\langle H_u^0 \rangle =  v_u  , \quad \langle H_d^0 \rangle =  v_d  .
\eeq
Note that our assignment of quantum numbers 
for the scalar and fermion multiplets 
as well as the form of the two doublet Higgs fields 
differ from those in Ref. \cite{Kaplan03} 
due to the different choice of the fermion multiplets. 
The two Higgs doublets in Eq. (\ref{Eq:Higgsfield}) are similar to 
those in the minimal supersymmetric model (MSSM) \cite{Djouadi09} 
since those could also be regarded as the pseudo Goldstone boson solution 
to the doublet-triplet splitting problem of 
supersymmetric grand unified theories (SUSY GUTs) \cite{Inoue86}.

  The Higgs vacua introduced above give masses to the SM fermions after the electroweak symmetry breaking (EWSM).  The corresponding lepton mass terms are identical in both the universal and anomaly-free embedding, and given by
\beq
\mathcal{L}^\ell =  \left(\lambda_{m1}^\ell i\ell_{m1}^c\Phi_{3}^\dagger + \lambda_{m2}^\ell  i\ell_{m2}^c\Phi_{4}^\dagger \right)\psi_m^{\ell} + \textrm{h.c} ,
\eeq
 where $\ell^{c}_{m1}$ and $\ell^{c}_{m2}$ are linear combinations of $e^c_m$ and $E^c_m$, and we neglect neutrino masses and mixings.  The above Lagrangian is invariant under $U(1)_X$.
As mentioned earlier, one can introduce one or more Weyl singlet states which may implement the appropriate neutrino oscillations and masses.  But we do not consider that possibility here since it is beyond the scope of this paper.  After $SU(4)_L$ is broken down to $SU(2)_L$, the $f$ vevs in Eq. (\ref{Eq:Scalarfield}) generate masses $m_{E_m}$ for $E_m$, and the orthogonal states $e_m$ remain massless:
{\setlength\arraycolsep{2pt} \beqr
E_m^c &=& (\lambda_{m1}^\ell f_3\ell_{m1}^c + \lambda_{m2}^\ell f_4\ell_{m2}^c)/m_{E_m}, \quad m_{E_m} = \sqrt{(\lambda_{m1}^\ell f_3)^2 + (\lambda_{m2}^\ell f_4)^2} , \cr
e_m^c &=& (-\lambda_{m2}^\ell f_4\ell_{m1}^c + \lambda_{m1}^\ell f_3\ell_{m2}^c)/m_{E_m} .
\eeqr }
After EWSB, the $v$ vevs in Eq. (\ref{Eq:Higgsvev}) produce small mass mixing in the lepton sector, and the lepton mass terms become
\beq
\mathcal{L}^\ell_{\textrm{mass}} = - \frac{\lambda_{m1}^\ell\lambda_{m2}^\ell f_{34}v_d}{\sqrt{2}m_{E_m}}e_m^c e_m
+ \frac{\left[(\lambda_{m1}^\ell)^2-(\lambda_{m2}^\ell)^2\right]f_3f_4v_d}{\sqrt{2}f_{34}m_{E_m}}E_m^c e_m - m_{E_m}E_m^cE_m
 + \textrm{h.c.}.
\eeq
After the lepton mass matrix is diagonalized, the SM leptons have the following masses
\beq
m_{e_m} = \frac{\lambda_{m1}^\ell\lambda_{m2}^\ell f_{34}v_d}{\sqrt{2}m_{E_m}}
 - m_{E_m}\xi_{e_m}^2 + O(\frac{1}{f^2}),
\eeq
where $\xi_{e_m}$ is a mixing angle between $e_m$ and $E_m$ defined by
\beq
\xi_{e_m} = \frac{\left[(\lambda_{m1}^\ell)^2-(\lambda_{m2}^\ell)^2\right]f_3f_4v_d}{2\sqrt{2}f_{34}m_{E_m}^2}.
\eeq
Note that both of $\lambda_{mi}^\ell$ must have non-zero values in order to have lepton masses positive.  If there is no mixing ($\lambda_{m1}^\ell = \lambda_{m2}^\ell$), the mass ratios of the charged SM leptons to the new heavy partners are simply given by the ratios of the vevs as $m_{e_m}/m_{E_M} = v_d/(\sqrt{2}f_{34})$.  Then, a few TeV of $f$ results in the existence of the heavy electron partner $E_1$ with a tenth of MeV mass, which is not plausible.  Therefore, the mixing terms proportional to the angle $\xi$ may be about the size of (but not exceed) the bare mass enough to have large hierarchy between $e_m$ and $E_m$ masses.

\begin{figure}[!h]
\centering%
  \subfigure[]{\label{Toploop1} %
    \includegraphics[height=2.5cm]{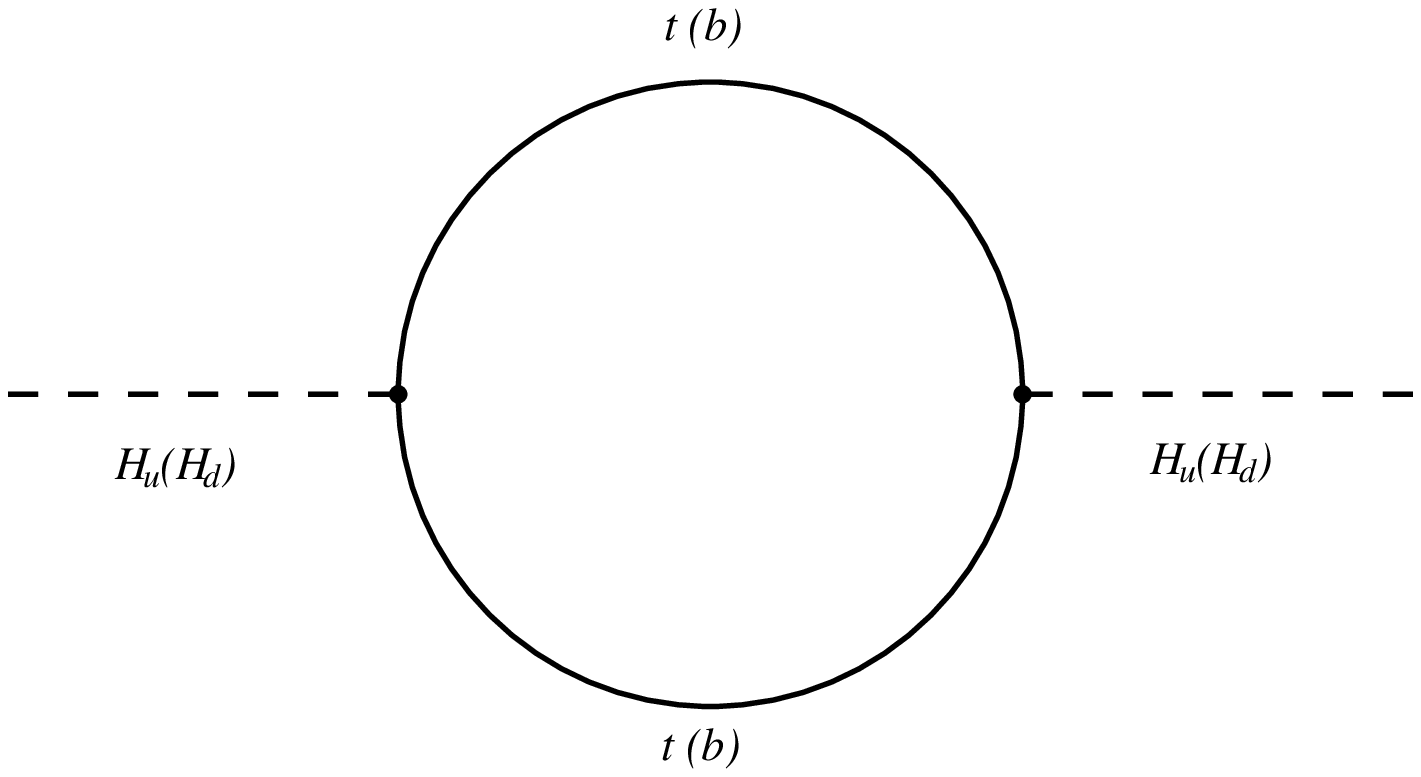}} \quad
  \subfigure[]{\label{Toploop2} %
    \includegraphics[height=2.5cm]{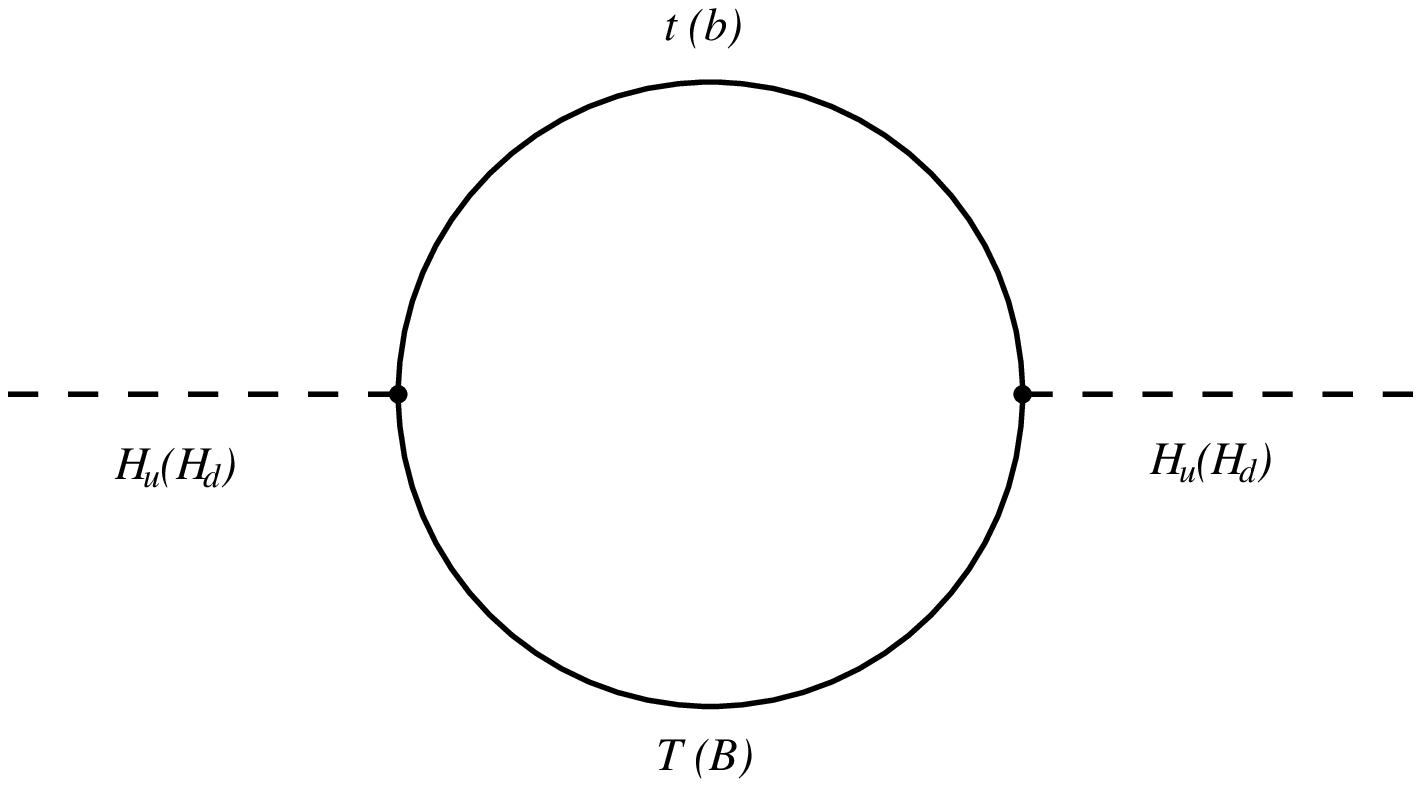}} \qquad
  \subfigure[]{\label{Toploop3} %
    \includegraphics[height=2.5cm]{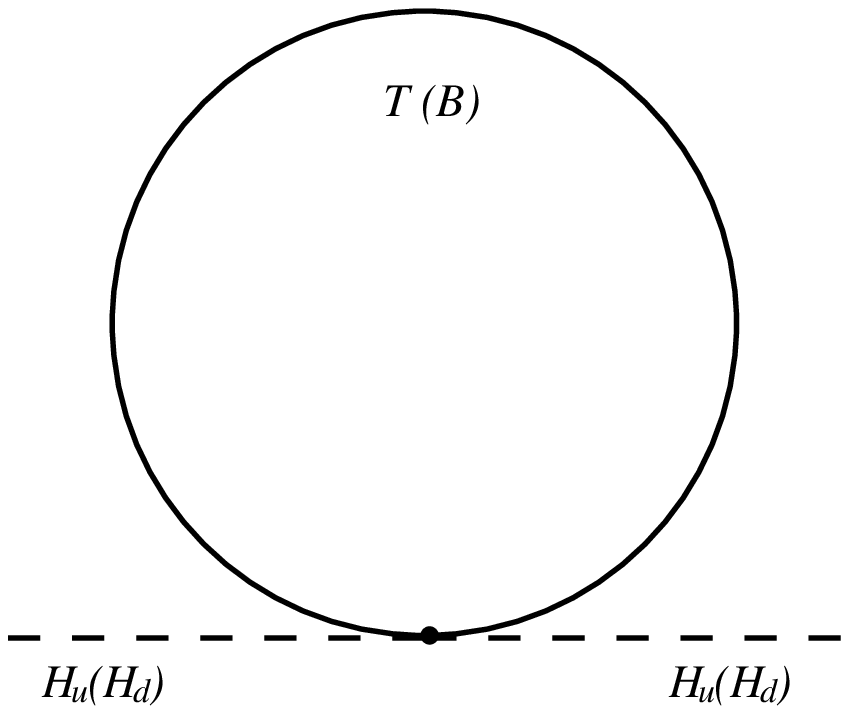}}
\caption{One-loop contributions to the Higgs mass from the top (bottom) sector.} \label{fig:Toploop}
\end{figure}

 The quark Yukawa Lagrangian in the universal embedding is analogous to that of the charged leptons.  But, it is more complicated in the anomaly-free embedding because of the different charge assignments for the first two generations of quark quadruplets.  The relevant $U(1)_X$ invariant quark mass terms are given by
{\setlength\arraycolsep{2pt} \beqr
\mathcal{L}_3^q &=& \left(\lambda_1^t iu_1^c\Phi_{1}^\dagger + \lambda_2^t iu_2^c\Phi_{2}^\dagger +
\lambda_1^b id_1^c\Phi_{3}^\dagger + \lambda_2^b id_2^c\Phi_{4}^\dagger \right)\psi_3^{q} + \textrm{h.c}. \cr
\mathcal{L}_n^q &=& \left(\lambda_{n1}^{u} iu_{n1}^{c}\Phi_{3}^\dagger + \lambda_{n2}^{u} iu_{n2}^{c}\Phi_{4}^\dagger +
\lambda_{n1}^{d} id_{n1}^{c}\Phi_{1}^\dagger + \lambda_{n2}^{d} id_{n2}^{c}\Phi_{2}^\dagger \right)\psi_n^{q} + \textrm{h.c}. ,
\eeqr }
where $n=1,2$; $u_{1,2}^c$ ($d_{1,2}^c$) are linear combinations of $t^c$ ($b^c$) and $T^c$ ($B^c$), and $u_{n1,n2}^c$ ($d_{n1,n2}^c$) are linear combinations of $u^c$ ($d^c$) and $U^c$ ($D^c$) for $n=1$ and of $c^c$ ($s^c$) and $C^c$ ($S^c$) for $n=2$.  Unlike the K-S model, $\Phi_{1,2}$ are responsible for the $up(down)$-type quark masses and $\Phi_{3,4}$ are for the $down(up)$-type quark masses of the 3rd family (first two families) because the $X$-charge of $\Phi_{1,2}$ is opposite to that of of $\Phi_{3,4}$, which results in the cancellation of one-loop quadratic divergences between the SM particles and their heavy partners of the \textsl{same} electric charges.
Before we obtain the quark masses, we first check the cancellation of quadratic divergences from top and bottom loops.  The couplings of the neutral Higgs to the $t (b)$ and $T (B)$ quarks are obtained from the above Lagrangian by expanding the nonlinear sigma fields $\Phi_i$:
{\setlength\arraycolsep{2pt} \beqr \label{eq:Higgsint}
\mathcal{L}_3^q &\supset& \lambda_tH^0_ut^ct + \lambda_{tT}H^0_uT^ct + \frac{\lambda_T}{2m_T}H^0_uH^0_uT^cT  \cr
 && + \lambda_bH^0_db^cb + \lambda_{bB}H^0_dB^cb + \frac{\lambda_B}{2m_B}H^0_dH^0_dB^cB + \textrm{h.c}. ,
\eeqr }
where
\bdpm
\lambda_t = \frac{\lambda_1^t\lambda_2^tf_{12}}{\sqrt{2}m_T} , \quad
\lambda_{tT} = \frac{\left[(\lambda_1^{t})^2-(\lambda_2^{t})^2\right]f_1f_2}{\sqrt{2}f_{12}m_T} , \quad
\lambda_T = \frac{(\lambda_1^tf_2)^2 + (\lambda_2^tf_1)^2}{2f_{12}^2} ,
\edpm
\beq \label{eq:lambdacouplings}
\lambda_b = \frac{\lambda_1^b\lambda_2^bf_{34}}{\sqrt{2}m_B} , \quad
\lambda_{bB} = \frac{\left[(\lambda_1^{b})^2-(\lambda_2^{b})^2\right]f_3f_4}{\sqrt{2}f_{34}m_B} , \quad
\lambda_B = \frac{(\lambda_1^bf_4)^2 + (\lambda_2^bf_3)^2}{2f_{34}^2} ,
\eeq
and where $m_T$ and $m_B$ are the masses of the heavy quark states $T$ and $B$, respectively:
\beq
m_T = \sqrt{(\lambda_1^{t}f_1)^2 + (\lambda_2^{t}f_2)^2} , \quad
m_B = \sqrt{(\lambda_1^{b}f_3)^2 + (\lambda_2^{b}f_4)^2} .
\eeq
The interaction terms in Eq. (\ref{eq:Higgsint}) lead to three diagrams shown in Fig. \ref{fig:Toploop} contributing to the Higgs masses.  The fact that one-loop quadratic divergences from the three diagrams in Fig. \ref{fig:Toploop} should be canceled in the LHMs demands the following condition for the couplings shown in Eq. (\ref{eq:Higgsint}):
\beq
\lambda_T - \lambda_t^2 - \lambda_{tT}^2 = 0 = \lambda_B - \lambda_b^2 - \lambda_{bB}^2 .
\eeq
One can easily check the validity of this condition using the relationship among the couplings shown in Eq. (\ref{eq:lambdacouplings}).  The very same cancellation procedure for top-quark loop was already discussed in the LHM with $SU(3)_L \times U(1)_X$ gauge group \cite{Han06}, but we extend this cancellation requirement to the bottom-quark sector as well which is commonly ignored in other type of LHMs including the K-S model.  Of cause, bottom loop contribution to the Higgs mass may not be crucial to resolve the fine-tuning issue, but it is important for the consistency of the model itself.  Similarly, one can confirm the cancellation of one-loop quadratic divergences from all other quarks and leptons.


 Similarly to the charged leptons after EWSM, the SM quarks have the following masses
\beq
m_t = \lambda_t v_u  - m_T\xi_t^2 + O(\frac{1}{f^2}), \quad
m_b = \lambda_b v_d  - m_B\xi_b^2 + O(\frac{1}{f^2}) ,
\eeq
where $\xi_t (\xi_b)$ is a mixing angle between $t (b)$ and $T (B)$ defined by
\beq \label{eq:quarkmixing}
\xi_t = \lambda_{tT}\frac{v_u}{2m_T} , \quad \xi_b = \lambda_{bB}\frac{v_d}{2m_B} .
\eeq
As discussed in the lepton case, the above quark mixings may not be negligible.
The masses of the remaining quarks in the first two generations can be obtained similarly.

\subsection{Gauge Boson Sector}

Since the quantum numbers of all scalar and fermion fields are identified, the gauge boson structure of the electroweak sector is fixed, and the 15 gauge fields $A_\alpha^\mu$ associated with $SU(4)_L$ can be written as
\beq \label{eq:GaugeMatrix}
T_\alpha A_\alpha =\frac{1}{\sqrt{2}}\left(
\begin{array}{cccc}D^{0}_{1} & W^{+} & Y^{0} & X^{\prime +}\\
W^{-} & D^{0}_{2} &  X^{-}_{} &  Y^{\prime 0}\\
\bar{Y}^{0} & X^{+} & D'^{0}_{1} & W^{\prime +}\\
X^{\prime -} & \bar{Y}^{\prime 0} & W^{\prime -} & D'^{0}_{2} \end{array}\right),
\eeq
where
$D_{1}^0=A_{3}/\sqrt{2}+A_{8}/\sqrt{6}+A_{15}/\sqrt{12}$,
$D_{2}^0=-A_{3}/\sqrt{2}+A_{8}/\sqrt{6}+A_{15}/\sqrt{12}$,
$D_{1}'^0=-2A_{8}/\sqrt{6}+A_{15}/\sqrt{12}$,
$D_{2}'^0=-3 A_{15}/\sqrt{12}$, and Lorentz index $\mu$ is omitted.
In order to check the cancellation of quadratic divergences from gauge boson loops, we first obtain the quartic couplings of the neutral Higgs to the gauge bosons from the scalar kinetic terms with the covariant derivative as follows:
{\setlength\arraycolsep{2pt} \beqr \label{eq:HiggsGauge}
\mathcal{L}_{kinetic}^{\Phi} &\supset&  \frac{1}{4}g^2\Big[ H_u^0H_u^0 \left( W^+W^- - W'^+W'^- - X^+X^- + X'^+X'^- + Z_1^2 - Z_1'^2 \right)   \cr
&& +  H_d^0H_d^0 \left( W^+W^- - W'^+W'^- + X^+X^- - X'^+X'^- + Z_2^2 - Z_2'^2\right)  \Big] ,
\eeqr }
where $Z_i^{(\prime)}$ is a linear combination of $D_i^{(\prime)}$ and $A_X$:
\beq
Z_1^{(\prime)} = D_1^{(\prime)} - \frac{g_X}{\sqrt{2}g}A_X , \quad
Z_2^{(\prime)} = D_2^{(\prime)} + \frac{g_X}{\sqrt{2}g}A_X .
\eeq
The quartic couplings in Eq. (\ref{eq:HiggsGauge}) lead to one-loop diagrams contributing to the Higgs masses similar to Fig. \ref{Toploop3}, and one can easily see from Eq. (\ref{eq:HiggsGauge}) that the cancellation of the quadratic divergences in the gauge boson sector is made between the primed and non-primed gauge bosons while one cannot expect such a cancellation in other $SU(4)_L \times U(1)_X$ models with usual Higgs scalars in which all Higgs-gauge boson couplings have same sign.

 After EWSB, the charged gauge bosons have following masses:
\bdpm
M^2_{W} = \frac{1}{4}g^2v^2, \quad M^2_{W^{\prime}} = \frac{1}{4}g^2\left(4f^2 - v^2 \right),
\edpm
\bdpm
M^2_{X} = \frac{1}{4}g^2\left(2f^2 - v_1^2 + v_2^2 \right),\quad M^2_{Y} = \frac{1}{2}g^2f^2,
\edpm
\beq
M^2_{X^\prime} = \frac{1}{4}g^2\left(2f^2 + v_1^2 - v_2^2 \right),\quad M^2_{Y^\prime} = \frac{1}{2}g^2f^2,
\label{Wmass}
\eeq
where
\beq
v_1^2 = v_u^2 - \dfrac{v_u^4}{3f^2}\left(\dfrac{f_{2}^2}{f_{1}^2}+ \dfrac{f_{1}^2}{f_{2}^2}-1\right), \quad
v_2^2 = v_d^2 - \dfrac{v_d^4}{3f^2}\left(\dfrac{f_{4}^2}{f_{3}^2}+ \dfrac{f_{3}^2}{f_{4}^2}-1\right), \quad
v^2 = v_1^2+v_2^2 ,
\eeq
and we use the following simplifying assumption $f_{12} = f_{34} = f$ as done in Ref. \cite{Kaplan03, Csaki03} for clear comparison. The three neutral gauge bosons $A^3$, $A^8$ and $A^{15}$ mixing with the $U(1)_X$ gauge boson $A^x$ are associated with a $4\times 4$ nondiagonal mass matrix. After the mass matrix is diagonalized, a zero eigenvalue corresponds to the photon $A$, and the three physical neutral gauge bosons $Z$, $Z'$ and $Z''$ have the following masses (squared):
\beq \label{Zmass}
M^2_Z = \frac{g^2v^2}{4c_W^2}\left(1 - \frac{t_W^4}{4}\frac{v^2}{f^2}\right), \quad
M^2_{Z'} = (g^2+g_X^2)f^2 - M^2_{Z}, \quad
M^2_{Z''} = \frac{1}{2}g^2f^2 ,
\eeq
where $c_W \equiv \cos{\theta_W} = \sqrt{(g^2+g_X^2)/(g^2+2g_X^2)}$, and $\theta_W$ is the Weinberg mixing angle.  Note that the extra neutral gauge boson $Z''$ does not mix with $Z$ or $Z'$ for $f_{12} = f_{34} = f$, but it still couples to ordinary fermions while $Z''$ in the K-S model only couples to the new heavy fermions.

The off-diagonal elements of the gauge boson matrix representation in Eq. (\ref{eq:GaugeMatrix}) do not mix each other, so $W^\pm$ are decoupled from other heavy bosons.
For the charged current, the fermion-gauge boson interaction terms (here we only show the SM fermions in one family) are then given by
{\setlength\arraycolsep{2pt} \beqr
\mathcal{L}_{CC} &=& -\frac{g}{\sqrt{2}}\left[\bar{t}\gamma^\mu(1-\gamma_5)b + \bar{\nu}\gamma^\mu(1-\gamma_5)e\right]W^+_\mu \cr
&&+ (\textrm{terms with}\ X^{(\prime)}, Y^{(\prime)}, \textrm{and}\ W^\prime) + \textrm{h.c}.,
\eeqr }
where other heavy bosons ($X^{(\prime)}, Y^{(\prime)}, W^\prime$) change flavors of new heavy fermions, and we do not consider their interactions here since we assume that the new fermions are too heavy to be seen below a few TeV.  On the other hand, the new neutral gauge bosons couple to the SM fermions directly or through mixing.  The neutral current is given by
{\setlength\arraycolsep{2pt} \beqr \label{eq:Ncurrents}
\mathcal{L}_{NC} &=& - e \mathnormal{Q}\left(\bar{\psi}\gamma^\mu \psi\right)A_\mu - \frac{g}{2c_W}\left[\bar{\psi}\gamma^\mu\left(g_V - g_A\gamma_5\right) \psi\right]Z_\mu
 \cr
&&  - \frac{g}{2c_W}\left[\bar{\psi}\gamma^\mu\left(g^{\prime}_V - g^{\prime}_A\gamma_5\right) \psi\right]Z^{\prime}_\mu + \frac{g}{4\sqrt{2}}\left[\bar{\psi}\gamma^\mu\left(1-\gamma_5\right) \psi\right]Z_\mu^{\prime\prime},
\eeqr }
where the values of $g_V^{(\prime)}$ and $g_A^{(\prime)}$ for the SM fermions are listed in Table \ref{tab:couplings}.\footnote{The contributions of the flavor mixings between the SM and the new heavy fermions to $g_V^{(\prime)}$ and $g_A^{(\prime)}$ are suppressed by $1/f^2$, so we neglect such flavor mixing effects.} 
The couplings in the table are family-universal unlike those in the K-S (Type I) model with anomaly-free fermion embbeding.
One can see from the table that the couplings contain additional new physics (NP) contributions proportional to the mixing angle $\theta$ between $Z$ and $Z'$ where $s_\theta \equiv \sin\theta =  t_W^2\sqrt{1-t_W^2}v^2/(2c_Wf^2)$.
Note that the masses of all heavy gauge bosons and the mixing angle $\theta$ are uniquely determined by the single parameter $f$.  We obtain the bounds of this paramter $f$ using the various electroweak precision data in the next section.

\begin{table}[hbt!]
\begin{ruledtabular}
\begin{tabular}{lllll}
  $\psi$ & $g_V$ & $g_A$ & $g_V^\prime$ & $g_A^\prime$ \\
  \hline
  $t$ & $\frac{1}{2}-\frac{4}{3}s_W^2 +\frac{5}{6}r s_Ws_\theta$
      & $\frac{1}{2} -\frac{1}{2}rs_Ws_\theta $
      & $\left(\frac{1}{2}-\frac{4}{3}s_W^2\right)s_\theta - \frac{5}{6}rs_W$
      & $\frac{1}{2}s_\theta + \frac{1}{2}rs_W $ \\
  $b$ & $-\frac{1}{2}+\frac{2}{3}s_W^2 -\frac{1}{6}rs_Ws_\theta$
      & $-\frac{1}{2} +\frac{1}{2}rs_Ws_\theta $
      & $\left(-\frac{1}{2}+\frac{2}{3}s_W^2\right)s_\theta +\frac{1}{6}rs_W$
      & $-\frac{1}{2}s_\theta - \frac{1}{2}rs_W $ \\
$\nu$ & $\frac{1}{2} -\frac{1}{2}rs_Ws_\theta$
      & $\frac{1}{2} -\frac{1}{2}rs_Ws_\theta $
      & $\frac{1}{2}s_\theta +\frac{1}{2}rs_W$
      & $\frac{1}{2}s_\theta + \frac{1}{2}rs_W $ \\
  $e$ & $-\frac{1}{2} +2s_W^2 -\frac{3}{2}rs_Ws_\theta$
      & $-\frac{1}{2} +\frac{1}{2}rs_Ws_\theta $
      & $\left(-\frac{1}{2} + 2s_W^2\right)s_\theta +\frac{3}{2}rs_W$
      & $-\frac{1}{2}s_\theta - \frac{1}{2}rs_W $
\end{tabular}
\caption{$Z$ and $Z'$ couplings to the SM fermions where $r \equiv g_X/g$.}
\label{tab:couplings}
\end{ruledtabular}
\end{table}

\section{Electroweak Constraints}

Electroweak precision test with the $Z$-pole data at LEP and SLC has provided highly accurate examination of the SM
and stringent constraints on the new physics beyond the SM \cite{LEPEWWG}.
Here we make the nonstandard precision analysis of our model with the precision variables $\epsilon_i$'s defined by Altarelli et al.
\cite{epsilon,altarelli}.
Our method provides a model independent way for the electroweak precision test.
Since the electroweak radiative corrections containing whole $m_t$ and $m_H$ dependencies
are parametrized in terms of $\epsilon_i$'s, the $\epsilon$'s can be extracted from the data
without specifying $m_t$ and $m_H$. We perform the modified analysis introduced in Ref. \cite{kylee1,kylee2},
in which the set of four variables $(\epsilon_1, \epsilon_3, \epsilon_b, \epsilon'_b)$
is defined by one-to-one correspondence to the observables $(\Gamma_l, A_{FB}^l, \Gamma_b, A_{FB}^b)$.
The variables $\epsilon_1$ and $\epsilon_3$ are extracted from the leptonic decay width $\Gamma_l$ and forward-backward
asymmetry $A_{FB}^l$. 
The variable $\epsilon_b$ is introduced to measure the additional contribution to the $Z b \bar{b}$ vertex
due to the large $m_t$-dependent corrections in the SM.  The variable $\epsilon'_b$ is required to measure
only the new physics effects of $Zb\bar{b}$ vertex and vanishes in the SM limit.
We assume that the new physics contributions to the $Z$-pole observables are comparable with the loop contributions of the SM.

We express the observable set in terms of the $\epsilon_i$'s up to the linear order given in Ref. \cite{altarelli,kylee1},
{\setlength\arraycolsep{2pt} \beqr
\Gamma_l &=& \Gamma_l |_B (1 + 1.20 \epsilon_1 - 0.26 \epsilon_3),
\nonumber \\
A_{FB}^l &=& A_{FB}^l |_B (1 + 34.72 \epsilon_1 - 45.15 \epsilon_3),
\nonumber \\
\Gamma_b &=& \Gamma_b |_B (1 + 1.42 \epsilon_1 - 0.54 \epsilon_3
+ 2.29 \epsilon_b -1.89 \epsilon'_b),
\nonumber \\
A_{FB}^b &=& A_{FB}^b |_B (1 + 17.50 \epsilon_1 - 22.75 \epsilon_3
+ 0.157 \epsilon_b -1.20 \epsilon'_b),
\eeqr }
where $\Gamma |_B$ and $A_{FB} |_B$ denote Born values defined by the tree level results including pure QED and QCD corrections.
We use the numerical values $\alpha_s(m_Z^2)=0.119$ and $\alpha(m_Z^2)=1/128.90$.
From the experimental values of $(\Gamma_l, A_{FB}^l, \Gamma_b, A_{FB}^b)$
in Ref. \cite{LEPEWWG},
we obtain $\epsilon_i$'s as\footnote{The obtained experimental value of $\epsilon_b$ does not agree with that in Ref. \cite{LEPEWWG} because they did not consider the $\epsilon'_b$ parameter.}
\beqr \label{eq:epsilon}
\epsilon_1 = (5.1\pm1.1) \times 10^{-3}, &&
\epsilon_3 = (3.8\pm1.8) \times 10^{-3},
\nonumber \\
\epsilon_b = (2.8\pm2.9) \times 10^{-2}, &&
\epsilon'_b = (3.5\pm4.0) \times 10^{-2},
\eeqr
which is a 4-dimensional ellipsoid in
$(\epsilon_1, \epsilon_3, \epsilon_b, \epsilon'_b)$ space.

We write the $\epsilon_i$ variables as $\epsilon_i = \epsilon_i^\textrm{SM}+\epsilon_i^\textrm{new}$, and use the following SM prediction $\epsilon_i^\textrm{SM}$ calculated by ZFITTER \cite{Bardin92}:
\beqr
\epsilon_1^\textrm{SM} = 5.8 \times 10^{-3}, &&
\epsilon_3^\textrm{SM} = 5.0 \times 10^{-3}, \nonumber \\
\epsilon_b^\textrm{SM} = -6.6 \times 10^{-3}, &&
\epsilon_b^{\prime\textrm{SM}} = 0.
\eeqr
The new physics contributions $\epsilon_i^\textrm{new}$ are extracted from
the vector and axial vector couplings of leptons and $b$ quark
to $Z$ boson presented in Table \ref{tab:couplings}:
\beqr
\epsilon_1^{\rm new} = -2r s_W s_\theta , ~~
\epsilon_3^{\rm new} = -r \frac{c_W^2}{s_W}s_\theta , ~~
\epsilon_b^{\rm new} = 0, ~~
{\epsilon'_b}^{\rm new} = -\frac{2}{3}r s_W c_W^2 s_\theta .
\eeqr
Using the experimental ellipsoid given in Eq. (\ref{eq:epsilon}),
we plot $\chi^2$ as a function of the mass of $Z'$ boson in Fig. 2.
We require $\chi^2 < 9.49$ for four degrees of freedom at 95\% CL
to yield the lower bound of $m_{Z'} > 1.1$ TeV.  As one can see from the figure, $\chi^2$ becomes minimum at around $m_{Z'} = 2.2$ TeV which corresponds to $f = 2.8$ TeV and $m_{Z''} = 1.3$ TeV.

\begin{figure}[!hbt]
\centering%
\includegraphics[width=9cm]{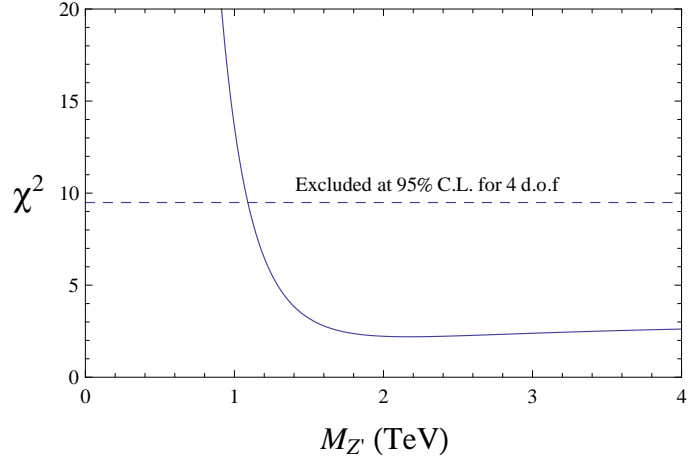}
\caption{$\chi^2$ plot for the observables $(\epsilon_1, \epsilon_2, \epsilon_b,\epsilon'_b)$ as a function of $Z'$ boson mass.
The region above the dashed line for $\chi^2 = 9.49$ is excluded with 95 \% C. L. for four degrees of freedom. }
\end{figure}

 Besides the above $\epsilon_i$ parameters, NP effects on precision measurements can be described by  the oblique parameter $\rho_0$ defined by $\rho_0 \equiv m_W^2/(m_Z^2 c_W^2 \hat{\rho})$
where $ \hat{\rho}$ includes the quadratic $m_t$ dependence, $ \hat{\rho} =1.01023 \pm 0.00022$.   
In the SM, $\rho_0=1$ exactly and our model expectation to the custodial $SU(2)_L$ symmetry violating shift $\Delta \rho_0 (\equiv \rho_0-1)$  in the $Z$ mass can be obtained from Eq. (\ref{Zmass}) by
\beq
\Delta \rho_0 = \alpha T \simeq \frac{g_X^4}{4 (g^2+g_X^2)^2} \frac{v^2}{f^2} ,
\eeq
where $T$ is the Peskin-Takeuchi parameter which has Higgs mass ($M_H$) dependance.
From the grobal fit, $\rho_0 = 1.0004^{+0.0027}_{-0.0007}$ at 2 $\sigma$ level with no meaningful bound on the Higgs mass \cite{pdg}.  Its error bar gives a bound of $f \geq 665$ GeV  which implies  $m_{Z'}\geq 511$ GeV, and its central value corresponds to $f = 1.85$ TeV and $m_{Z'} = 1.44$ TeV.  On the other hand, if we assume that a lighter CP-even Higgs in this model is SM-like one,  one has $T \leq 0.10 (0.19)$ at 95$\%$ CL for $M_H = 117$ GeV (300 GeV) which implies $f \geq 1.3$ TeV (960 GeV) and $m_{Z'} \geq 1.0$ TeV (750 GeV).

The masses of two additional neutral gauge bosons $Z'$ and $Z''$ can be constrained by a various low-energy experiments such as atomic parity violation (APV).   APV is sensitive to electron-quark interactions describing neutral current interaction processes written as
\beq
H_{\rm eff} = -\frac{G_F}{\sqrt{2}} \sum_{q=u,d}
       \left( C_{1q}~ \bar{e} \gamma^\mu \gamma_5 e~ \bar{q} \gamma_\mu q
     +  C_{2q}~ \bar{e} \gamma^\mu e~ \bar{q} \gamma_\mu \gamma_5 q \right),
\eeq
where the second term is strongly suppressed because of its dependence on spins rather than charges and the smaller vector coupling of electrons \cite{Altarelli91}.  The above interactions get contributions from $Z^{(\prime)}$ and $Z''$ exchanges expressed in Eq. (\ref{eq:Ncurrents}), and the relevant experimental results are  represented by the weak charge of an atom defined by
\beq
Q_W \equiv -2 \left[ C_{1u} (2Z+N) + C_{1d} (Z+2N) \right],
\eeq
where $Z(N)$ is the number of protons (neutrons) of the atom and the coefficients $C_{1q}$ in our model are given by
\beq
C_{1q} =  2 g_A^e g_V^q + 2 {g'}^e_A {g'}^q_V \frac{m_Z^2}{m_{Z'}^2} + \frac{1}{4} \cos^2 \theta_W \frac{m_Z^2}{m_{Z''}^2}.
\eeq
For the Cesium atom with $Z=55$ and $N=78$, the most recent average of weak charge measurements, $Q^\textrm{exp}_W=-73.16\pm 0.35$, is in good agreement with the SM prediction, $Q^\textrm{SM}_W=-73.16 \pm 0.03$ \cite{Porsev09}.  This result leads to the following bound of the new parameters at 95\% CL: 
\beq \label{eq:APVbound}
f \geq 2.8\ \textrm{TeV}, \quad m_{Z'} \geq 2.2\ \textrm{TeV}, \quad m_{Z''} \geq 1.3\ \textrm{TeV}.
\eeq
APV provides stronger constraints on the model parameters 
than the $Z$-pole data 
because the contribution of the $Z''$ boson exchange dominates in $Q_W$ 
while $\epsilon_i$ are determined by the $Z$ boson couplings only.

\section{Conclusions}

In summary, we discussed the aspects of the $SU(4)_L \times U(1)_X$ models with little Higgs in the scalar sector as an alternative solution to the hierarchy and fine-tuning issues.  We introduced a set of fermions which ensure the cancellation of gauge anomaly, and
explicitly showed the cancellation of the one-loop quadratic divergences to the Higgs masses from all fermion multiplets and all gauge bosons.  The new charged (flavor-changing) gauge bosons do not mix with the SM gauge bosons so that this model does not receive strong electroweak constraints in the charged sector.  On the other hand, there exist two extra neutral (flavor- conserving) gauge bosons $Z'$ and $Z''$ of which effects could appear at the low energy scale through mixing and/or direct exchange.  Using the recent experimental data, we obtained the bounds on the NP scale parameter $f$ and the masses of $Z'$ and $Z''$.  Especially, APV experiments give a strongest constraint because the contribution of the $Z''$ boson exchange is much bigger that that of $Z'$ exchange or $Z-Z'$ mixing.

The search for these extra heavy bosons is an important issue
of the experimental program of current and future high-energy colliders 
in order to test the NP models. 
Since $Z''$ is the lightest new gauge boson in this model 
and its couplings to fermions are not very small 
while other charged heavy bosons do not couple to the SM fermion pairs, 
there is a possibility that it could be discovered at the Tevatron 
in the Drell-Yan process $p\bar{p} \to Z'' \to \ell^+\ell^-$ with $\ell =e,\mu$ 
if lighter than about 1 TeV.  
If the bound in Eq. (\ref{eq:APVbound}) is solidated in future experiments, 
however, one could instead expect its discovery at the LHC  
using the process  $p\bar{p} \to Z'' \to \ell^+\ell^-$ 
if lighter than about 5 TeV. 
A further detailed study on the collider signatures of $Z'$ and $Z''$ bosons 
as well as the charged Higgs scalars in this model is in progress.   

\begin{acknowledgments}
S.-h. Nam thanks Otto Kong for collaboration at the beginning of this project.
The work of K. Y. L. was supported by WCU program through the KOSEF
funded by the MEST (R31-2008-000-10057-0) and by the Korea Research Foundation Grant
funded by the Korean Government (KRF-2008-313-C00167).
Y.Y.K's work was partially supported by APCTP in Korea and
was supported in part by National Research Foundation of Korea Grant funded by the Korean Government 2009-0072611.

\end{acknowledgments}


\end{document}